\documentclass[12pt]{article}
\usepackage{amsmath,amssymb}
\usepackage{graphicx,color,xcolor}
\usepackage{framed,multirow}
\usepackage{paralist,url}
\colorlet{framecolor}{black}
\colorlet{shadecolor}{lightgray}
 \allowdisplaybreaks
\newcommand{\be}{\begin{equation}}
\newcommand{\ee}{\end{equation}}
\newcommand{\bea}{\begin{eqnarray}}
\newcommand{\eea}{\end{eqnarray}}

\newcommand{\tdg}{{\td g}}

\newcommand{\cA}{{\cal A}}

\newcommand{\cO}{{\cal O}}

\newcommand{\ds}{\displaystyle}
\newcommand{\nn}{\nonumber}
\newcommand{\pd}{\partial}
\newcommand{\td}{\tilde}

\long\def\symbolfootnote[#1]#2{\begingroup%
\def\thefootnote{\fnsymbol{footnote}}\footnote[#1]{#2}\endgroup}

\begin{document}

\thispagestyle{empty}
\begin{center}


{\Large\bf A note on the Kerr-Newman metric in five dimensions}

\vspace{10pt}

Hui-Min Fan, Zheng-Cheng Liang and Jianwei Mei\symbolfootnote[1]{Email:~\sf jwmei@hust.edu.cn}

\vspace{10pt}

{\it MOE Key Laboratory of Fundamental Physical Quantities Measurements,\\
School of Physics, Huazhong University of Science and Technology,\\
1037 Luoyu Rd., Wuhan 430074, P.R. China}

\vspace{65pt}
{\bf Abstract}
\end{center}
\vspace{-10pt}

We derive an ansatz for the five dimensional equal-rotation Kerr-Newman metric that contains two unknown functions. By solving for these functions through perturbation series, we find that the metric can be cast into the Kerr-Shild form in a background that is flat at the spatial infinity.

 \newpage

\tableofcontents

\section{Introduction}

Analytical black hole solutions are useful probes to the nature of gravitational field equations.\footnote{We are most interested in the nonlinearity of Einstein equations, which has been a main source of difficulty, and which is also a source of puzzle when schemes like the post-Newtonian approximation works unexpectedly well outside its realm of validity \cite{Will:2011nz}. There is expectation that experimental information on the strong field dynamics of gravity will become available from gravitational wave detectors such as the advanced LIGO and the advanced VIRGO within the next few years \cite{Aasi:2013wya}. Further information may also be available from proposed spaceborne detectors such as eLISA \cite{Seoane:2013qna} and TianQin \cite{Luo:2015ght} in the long run. This offers the real possibility to better understand the nonlinear nature of gravitational field equations, and it is important to study the problem even further from purely theoretical perspective. A related recent work can be found in \cite{Harte:2014ooa}.} A good example is a solution that can be cast into the Kerr-Schild form \cite{KS1965}, for which the metric is written as $\td{g}_{ \mu\nu} =g_{\mu \nu} +h_{\mu\nu}\,$, with\footnote{In this work, everything with a tilde is defined using $\td{g}_{\mu\nu}$, while all others are defined using $g_{\mu\nu}$.}
\be h_{\mu\nu}=f\xi_\mu \xi_\nu\,,\quad \xi^\mu \xi_\mu =\xi^\mu\nabla_\mu \xi_\nu=0\,,\label{KS}\ee
where $f$ is an arbitrary function. In this case, the classical Einstein equation is equivalent to its linearized form, and as a consequence, the solution by itself only probes dynamics of the linearized Einstein gravity.

It is an intriguing fact that all the physically most important black holes in four dimensions can be written in the Kerr-Schild form and they obey linearized field equations. These include the Schwarzschild black hole, the Reissner-Nordstr\"{o}m black hole, the Kerr black hole and the Kerr-Newman black hole. For higher dimensions, the Reissner-Nordstr\"{o}m black hole has been generalized by Tangherlini in 1963 \cite{Tangherlini:1963bw} and the Kerr black hole by Myers and Perry in 1986 \cite{Myers:1986un}. In contrast, the five dimensional analogue of the Kerr-Newman black hole is still not known in a closed form, while some numerical result has been obtained previously \cite{Kunz:2005nm,Kunz:2006eh}.

Since both the Tangherlini solution and the Myers-Perry black holes can be cast into the Kerr-Shild form, it is a natural question to ask, as whether the same can be done for the five dimensional Kerr-Newman black hole. Without a full knowledge of the metric, it has been difficult to answer the question.

There have been previous attempts to construct the exact Kerr-Newman black hole in the five dimensional pure Einstein-Maxwell theory. Most useful for us is \cite{NavarroLerida:2007ez}, which, based on experience from \cite{Kunz:2005nm,Kunz:2006eh}, has studied a simple ansatz for the equal rotation case. A perturbation expansion around the small electric charge has been carried out to the subsubleading order in the extremal case. This was generalized to include a Chern-Simons term with an arbitrary coupling constant in \cite{Allahverdizadeh:2010xx}. The small rotation limit of the solution has been studied in \cite{Aliev:2005npa,Aliev:2006yk}. In six and higher dimensions, more clues of the Kerr-Newman-like solutions can also be obtained by using the blackfold method \cite{Caldarelli:2010xz}.

In this work, we try to move one step further towards finding the exact five dimensional Kerr-Newman black hole. By starting from the ansatz used in \cite{NavarroLerida:2007ez}, which contains six unknown functions in total, we have managed to partially integrate the equations of motion and have thus reduced the total number of unknown functions to two.

Unfortunately, finding the exact result for the two last unknown functions turns out to be extremely difficult. We have tried to solve the equations through perturbation expansion, both by expanding around large radius and around small electric charge, but without a final success. However, the result obtained so far does help us show that the metric can be cast into the Kerr-Schild form in a background that is flat at the spatial infinity. We use this to study the nonlinearity of the field equations to be obeyed by the full solution.

\section{The Ansatz}

Our starting point is the equal rotation ansatz used in \cite{NavarroLerida:2007ez},
\bea ds^2&=&-Fdt^2+\frac{dr^2}W +r^2\Big[\frac{dx^2}{1-x^2}+(1-x^2) d\phi_1^2 +x^2 d\phi_2^2\Big]\nn\\
&&+N\Big[(1-x^2)d\phi_1 +x^2 d\phi_2\Big]^2-2B\Big[(1-x^2)d\phi_1 +x^2 d\phi_2\Big]dt\,,\nn\\
\cA&=&A_tdt+A_\phi\Big[(1-x^2)d\phi_1 +x^2 d\phi_2\Big]\,, \label{ansatz}\eea
which has the remarkable feature that, all the unknown functions $F\,,\, W\,,\, N\,,\, B\,,\, A_t$ and $A_\phi$ only depend on $r\,$. A similar ansatz has been used in \cite{Allahverdizadeh:2010xx} to study charged rotating black holes in general odd dimensions, where the Einstein-Maxwell Lagrangian is supplemented with a Chern-Simons term with an arbitrary coupling constant.

The equations were originally solved in \cite{NavarroLerida:2007ez,Allahverdizadeh:2010xx} for extremal black holes, by expanding the many unknown functions in the limit of small electric charge (characterized by a parameter $q\,$, defined through $A_t\sim \cO(q)\,$). Results up to the cubic order in the $q$ expansion is known.

We find that some of the field equations can be integrated for the ansatz (\ref{ansatz}). This is done by firstly introducing a new function $Z\,$, such that
\be W=\frac{16Z(c_0+Z)}{r^2Z'^2}\,,\ee
where $c_0$ is some constant, and $f'\equiv\pd_rf\,,\;\forall\;f\,$. With this we find
\be N=\frac{Z}{r^2F}-\frac{B^2}{F}-r^2\,.\ee
The metric can now be written as
\bea ds^2&=&-F\Big(dt+\frac{B}{F}[(1-x^2)d\phi_1+x^2d\phi_2]\Big)^2 +\frac{r^2Z'^2dr^2}{16Z(c_0+Z)}\nn\\
&&+r^2\Big[\frac{dx^2}{1-x^2}+(1-x^2)d\phi_1^2+x^2d\phi_2^2\Big]\nn\\ &&+\Big(\frac{Z}{r^2F}-r^2\Big)[(1-x^2)d\phi_1+x^2d\phi_2]^2\,. \label{ansatz2}\eea

We further find that $A_t$ only appears in the equations through its first and second order derivatives. This leads to
\bea A_t&=&\int\frac{(2r^2B\sqrt{c_0+Z}\,A'_\phi+\sqrt3\, qZ')F}{2(r^2B^2-Z)\sqrt{c_0+Z}}\,dr\,,\nn\\
F&=&\frac{2(r^2B^2-Z)Z'\td{A}_\phi}{r^2(2Z\sqrt{c_0+Z}\, A'_\phi+\sqrt3\,qBZ')}\,,\eea
where $\td{A}_\phi$ is determined by
\be\td{A}_\phi=-\int\frac{A_\phi Z'}{4r^2\sqrt{c_0+Z}}\,dr \,.\ee
Finally, $B$ is found to be
\bea B&=&-\frac{2Z\sqrt{c_0+Z}\,A'_\phi}{\sqrt3\,qZ'}\nn\\ &&-\frac{2\td{A}_\phi}{3\sqrt3\,qZ'^3} \Big\{12(c_0+Z)(ZZ'-rZ'^2+rZZ'')-Z'^2(6rZ-3r^2Z'+A_\phi^2Z'^2)\nn\\
&&\qquad\qquad+\Big[\Big(12(c_0+Z)(ZZ'-rZ'^2+rZZ'')-Z'^2(6rZ-3r^2Z' +A_\phi^2Z'^2) \Big)^2\nn\\
&&\qquad\qquad+3Z'^4(4r^2Z(c_0+Z)A'^2_\phi-3q^2Z'^2)\Big]^{1/2} \Big\}\,.\eea
With these, everything has been determined through $Z$ and $A_\phi\,$. And we have thus reduced the total number of unknown functions to two.

Unfortunately, the equations for $Z$ and $A_\phi$ are very unwieldy. We have tried to solve the equations through perturbation series, both by expanding around large radius ($r\to\infty$) and around small electric charge ($q\to 0$). In this process, we have chosen $c_0=m^2 - 2 m a^2 - q^2$ and have picked the coordinates so that the solution describes a black hole with mass $M=\frac34\pi m\,$, angular momenta $J_1=J_2=\frac\pi2ma$ and electric charge $Q=\frac{\sqrt3}4\pi q\,$.

More details are in the following subsections.

\subsection{Expansion around large radius}

We have tried to expand $Z$ and $A_\phi$ around $r\to\infty\,$, up to dozens of orders, hoping to find a recognisable pattern in the coefficients. This has not been successful yet.

The first few terms in the expansion are
\bea Z&=&r^4\Big(1-\frac{2m}{r^2}+\frac{2a^2m+q^2}{r^4} +\frac{a^2q^2s^2}{3 r^6}-\frac{8a^2mq^2(1+s)-7a^2mq^2s^2}{15r^8}+\cdots\Big)\,,\nn\\
A_\phi&=&\frac{\sqrt3\,aq}{r^2}\Big[s+\frac{4m(1+s)}{3\,r^2} +\frac{2 m^2(1+s)}{r^4}-\frac{1}{45r^6}(48a^2m^2-144m^3\nn\\
&&+20mq^2+48a^2m^2s-144m^3s+20mq^2s+9a^2q^2s^3)+\cdots\Big]\,, \label{large.r}\eea
where $s$ is a free numerical constant, and the value of $q$ is fixed through
\be A_t=\frac{\sqrt3\,q}{r^2}+\cO\Big(\frac1{r^3}\Big)\,.\ee

At the spatial infinity, we find (in the notation of \cite{Aliev:2006yk})
\be\hat{B}_2=-\frac{2\sqrt3qas}{r^2}\Big[xdx\wedge(d\phi_1-d\phi_2) +\frac{ dr \wedge[(1-x^2)d\phi_1+x^2d\phi_2]}r\Big]+\cO(\frac1{r^4})\,,\ee
corresponding to a gyromagnetic ratio $g=-3s\,$.

\subsection{Expansion around small charge}

We have also tried to expand $Z$ and $A_\phi$ around $q\to0$ in the extremal case (which has to $m=2a^2\,$), just like in \cite{NavarroLerida:2007ez}. To find possible recognizable patterns, we have pushed the perturbation series a couple of orders higher than those in \cite{NavarroLerida:2007ez,Allahverdizadeh:2010xx}. The result is ($u\equiv\sqrt2\,a/r$)
\bea Z&=&\frac{4a^4(1-u^2)^2}{u^4} +q^2\Big(\frac{30-55u^2 +27u^4}{30(1-u^2)^2} +q^2z_1 +q^4z_2+q^6z_3 +\cdots\Big)\,,\nn\\
A_\phi&=&q\,\Big(-\frac{\sqrt3\,u^2}{2a}+q^2p_1+q^4p_2 +q^6p_3 +\cdots\Big)\,,\label{small.q}\eea
where the first few coefficient functions are given in the appendix \ref{sec.coef}.

In this process, we have set $s=-1\,$, and the results coincide with (\ref{large.r}) in regions where the two types of expansion overlap.

A different convention was used in \cite{NavarroLerida:2007ez} (there the black hole mass was found to be $M=\frac32\pi a^2+\frac{\pi q^2}{8a^2}+\cdots\,$) and so the corresponding expressions do not appear identical.

The appearance of Polylog functions $Li_n(u^2)$ in the coefficients at higher orders not only shows how complicated the full solution can be, but it has also made it very difficult to arrange the results into a fully recognisable form.

\section{The Kerr-Schild form}

By pushing the series expansion in (\ref{large.r}) up to the order $\cO(\ds\frac1{r^{30}})$ for both $\ds\frac{Z}{r^4}$ and $r^2A_\phi\,$, we have found some confidence to say that, with the redefinition
\bea t&\longrightarrow&t+F_1(r)dr\,,\nn\\
\phi_1&\longrightarrow&\phi_1+F_2(r)dr\,,\nn\\
\phi_2&\longrightarrow&\phi_2+F_2(r)dr\,,\eea
the metric (\ref{ansatz2}) can be cast into the form of $\td{g}_{ \mu\nu} =g_{\mu\nu}+f\xi_\mu \xi_\nu\,$, with
\bea\xi&=&dt+R_2dr+P_2[(1-x^2)d\phi_1+x^2d\phi_2]\,,\nn\\
f&=&\frac{F(Z-r^4F)}{r^2[r^2F+(B-c_\phi F)^2]-Z}\,,\nn\\
g_{\mu\nu}dx^\mu dx^\nu&=&-f_1\xi'^2+f_2dr^2 +r^2\Big[ \frac{dx^2}{1-x^2} +(1-x^2)d\phi_1^2 +x^2d\phi_2^2\Big]\,,\nn\\
\xi'&=&dt+R_1dr+P_1[(1-x^2)d\phi_1+x^2d\phi_2]\,,\nn\\
f_1&=&\frac{r^2F\,(B-c_\phi F)^2}{r^2[r^2F+(B-c_\phi F)^2]-Z}\,,\nn\\
f_2&=&\frac{r^2[r^2F+(B-c_\phi F)^2]-Z}{16r^4F(B-c_\phi F)^2 (c_0+Z)} Z'^2\,,\nn\\
P_1&=&\frac{r^2B^2+r^4F-c_\phi r^2BF-Z}{r^2F(B-c_\phi F)}\,,\quad
P_2=c_\phi\,,\nn\\
R_1&=&F_1-\frac{(r^4F-Z)[Z-r^2B(B-c_\phi F)]Z'\sqrt{Z-r^2(B-c_\phi F)^2}}{4r^5F^{3/2}(B-c_\phi F)^2 Z\sqrt{c_0+Z}}\;,\nn\\
R_2&=&F_1+\frac{(r^4B-c_\phi Z)Z'\sqrt{Z-r^2(B-c_\phi F)^2}}{4r^3 \sqrt{F}\,(B-c_\phi F)Z\sqrt{c_0+Z}} \;,\nn\\
F_2&=&\frac{(r^4F-Z)Z'\sqrt{Z-r^2(B-c_\phi F)^2}}{4r^3\sqrt{F}(B-c_\phi F)Z \sqrt{c_0+Z}}\,.\label{metric5d}\eea
Here $c_\phi$ is an arbitrary constant and $F_1(r)$ is a free function. Given that $F_1(r)\to0\,$ as $r\to\infty\,$, then the background metric $g_{\mu\nu}$ in (\ref{metric5d}) approaches that of a flat spacetime at the spatial infinity.

We find that $\xi^\mu$ obeys (\ref{KS}), showing that the five dimensional equal rotation Kerr-Newman metric (\ref{ansatz2}) can be cast into the Kerr-Schild form.

A full analogy with the four dimensional Kerr-Newman black hole would have required that $\cA\propto\xi\,$, but this is not true for (\ref{metric5d}): The $R_2\,dr$ term can always be included into $\cA$ after a gauge choice, but
\bea P_2=c_\phi&\neq&\frac{A_\phi}{A_t}=as+\frac{4am(1+s)}{3r^2}+\frac{2am (1+s)(3m+a^2s)}{3r^4}\\
&&+\frac{2a[2m(36m^2-5q^2)(1+s)+3q^2a^2s^3+m^2a^2(50s^2+46s-4)]}{45 r^6} +\cdots\,.\nn\eea
Setting $c_\phi=as\,$, one has $\cA\approx A_t\,\xi$ at the spatial infinity.

To see how these properties can affect the equations of motion, we briefly recall some basic properties of (\ref{KS}). The inverse metric is
\be\tdg^{\mu\nu}=g^{\mu\nu}-h^{\mu\nu}\,.\ee
The affine connection is
\be\td\Gamma^\rho_{\mu\nu}=\Gamma^\rho_{\mu\nu}+K^\rho_{\mu\nu}\,,\quad K^\rho_{\mu\nu}=\frac12\Big(\nabla_\mu h_\nu^\rho +\nabla_\nu h_\mu^\rho -\nabla^\rho h_{\mu\nu}+h^{\rho\sigma} \nabla_\sigma h_{\mu\nu}\Big)\,,\ee
which leads to $K^\rho_{\rho\nu}=0\,$. The Ricci tensor is \cite{Gibbons:2004uw}
\bea\td{R}_\mu^\nu&=&R_\mu^\nu-h^{\nu\rho}R_{\mu\rho} +\frac12\nabla_\rho \Big(\nabla_\mu h^{\nu\rho} +\nabla^\nu h_\mu^\rho -\nabla^\rho h_\mu^\nu \Big)\,.\label{Ricci}\eea

Now consider the Einstein-Maxwell system,
\be S=\int d^nx\sqrt{|\td{g}|}\;\Big(\td{R}-\frac14\td{F}_{\mu\nu} \td{F}^{\mu \nu} \Big)\,,\ee
where $\td{F}_{\mu\nu}=\nabla_\mu \cA_\nu-\nabla_\nu \cA_\mu$ and
\be \td{F}^{\mu \nu} =\td{g}^{\mu\rho} \td{g}^{\nu \sigma} \td{F}_{\rho \sigma} =(g^{\mu\rho}-h^{\mu\rho})(g^{\nu \sigma} -h^{\nu \sigma}) \td{F}_{\rho \sigma}\,.\ee
Note $\td{F}^{\mu \nu} =g^{\mu\rho} g^{\nu \sigma} \td{F}_{\rho \sigma}$ in the case when $\cA_\mu=const.\; \xi_\mu\,$, while otherwise there is correction at the order $\cO(h_{\mu\nu}^2)$.

The Maxwell equations are
\be\td\nabla_\mu\td{F}^{\mu\nu}=0\,,\label{eom.A}\ee
which, given (\ref{KS}), are equivalent to
\be \nabla_\mu\td{F}^{\mu\nu}=0\,.\ee
So the Maxwell equations are not perturbed in the case $\cA_\mu =const.\; \xi_\mu\,$, but there is correction at the order $\cO(h_{\mu\nu}^2)$ in other cases.

The Einstein equations are
\be \td{R}_\mu^\nu=\frac12\td{F}_{\mu\rho}\td{F}^{\nu\rho} -\frac{ \td{F}_{\alpha\beta} \td{F}^{\alpha\beta}}{4(n-2)} \delta_\mu^\nu\,. \label{eom.g}\ee
The left hand side is linear in $h_{\mu\nu}\,$, as is obvious from (\ref{Ricci}). On the right hand side, however, there is either no correction from $h_{\mu\nu}$ when $\cA_\mu =const.\; \xi_\mu\,$, or otherwise there is correction up to the order $\cO(h_{\mu\nu}^2)\,$.

As such, although the metric of the five dimensional equal rotation Kerr-Newman metric can be cast into the Kerr-Schild form, but because $\cA_\mu\neq const.\;\xi_\mu\,$, the equations still cannot be fully linearized due to $\cO(h_{\mu\nu}^2)$ corrections in the Maxwell sector.

\section{Summary}

We have derived an ansatz for the five dimensional Kerr-Newman metric with equal rotations. Through perturbation expansion, we find that the metric can be cast into the Kerr-Schild form in a background that is flat at the spatial infinity. The equations are not fully linearized due to the fact that $\cA_\mu\neq const.\; \xi_\mu\,$, in contrast to the case in four dimensions.

The appearance of nonrational functions in the expansion coefficients in the appendix \ref{sec.coef} indicates that the five dimensional Kerr-Newman metric is likely to offer more information about the underlying field equations than its four dimensional counterpart. So it is highly desirable to find the full analytical solution and to study the general case when the two rotations are independent.

\section*{Acknowledgement}

J.M. thanks Slava Didenko for pointing out a loophole in the original version of the paper. This work was supported by the National Natural Science Foundation of China (Grants No. 11475064).

\appendix

\section{Coefficients in the expansion around small charge}
\label{sec.coef}

\bea z_1&=&-\frac{1}{20160a^4u^4(1-u^2)^6}(10080-78960u^2+270480u^4 -529480u^6\nn\\
&&+648536u^8-509320u^{10}+250665u^{12}-70800u^{14}+8827u^{16})\nn\\
&&-\frac{(1-u^2)^2(3-u^2)\ln(1-u^2)}{6a^4u^6}\,,\nn\\
z_2&=&\frac{1}{399168000a^8u^4(1-u^2)^{10}}(52557120-626287200u^2\nn\\
&&+3416471520u^4-11264810480u^6+24995591544u^8-39364775912u^{10}\nn\\
&&+45250961492u^{12}-38467816464u^{14}+24225312760u^{16}-11190924899u^{18}\nn\\
&&+3694414502u^{20}-824691567u^{22}+110718760u^{24}-6814500u^{26}+115500u^{28})\nn\\
&&+\frac{(2844-10100u^2+11600u^4-1525u^6-4255u^8+1550u^{10})\ln(1-u^2)}{21600a^8u^6(1-u^2)^2}\nn\\
&&+\frac{(1-u^2)^2(41-21u^2)\ln(1-u^2)^2}{360a^8u^6}\nn\\
&&-\frac{u^2(5+u^2)}{180a^8(1-u^2)^2}Li_2(u^2)\,,\nn\\
z_3&=&-\frac{1}{52306974720000a^{12}u^4(1-u^2)^{14}}(17655103065600\nn\\
&&-278621953209600u^2+2060306528280000u^4-9474066455453699u^6\nn\\
&&+30311868293426720u^8-71513965929486080u^{10}+128606135609094000u^{12}\nn\\
&&-179650418255634040u^{14}+196740396935785560u^{16}\nn\\
&&-169150124307844680u^{18}+113469550204439520u^{20}\nn\\
&&-58492303079650440u^{22}+22519002754713170u^{24}\nn\\
&&-6140584039634630u^{26}+1057213612835255u^{28}-76106588777580u^{30}\nn\\
&&-7653664324031u^{32}+1797547752000u^{34}-84441357000u^{36}+2270268000u^{38})\nn\\
&&-\frac{1}{21772800a^{12}u^6(1-u^2)^6}(5647944-46989584u^2+172571416u^4\nn\\
&&-365938224u^6+490623218u^8-425838394u^{10}+232951721u^{12}-72066675u^{14}\nn\\
&&+7971327u^{16}+1414759u^{18}-346500u^{20}+6300u^{22})\ln(1-u^2)\nn\\
&&+\frac1{362880a^{12}u^8(1-u^2)^2}(28350-147059u2+303867u^4-310996u^6+142315u^8\nn\\
&&-15399u^{10}-1414759u^{12})\ln(1-u^2)^2\nn\\
&&+\frac{(1-u^2)^2(-18+13u^2)\ln(1-u^2)^3}{810a^{12}u^6}\nn\\
&&+\frac{Li_2(u^2)}{181440a^{12}u^4(1-u^2)^6}(15120-115080u^2+385560u^4\nn\\
&&-736960u^6+873992u^8-655760u^{10}+304531u^{12}
-81020u^{14}+9869u^{16})\nn\\
&&\frac{(45-185u^2+290u^4-230u^6+50u^8-12u^{10})\ln(1-u^2)Li_2(u^2)}{540a^{12} u^6(1-u^2)^2}\nn\\
&&-\frac{90-370u^2+595u^4-465u^6+130u^8-22u^{10}}{540a^12u^6(1-u^2)^2} \Big[Li_3(1-u^2)\nn\\
&&\qquad\qquad\qquad\qquad\qquad+\ln(u)\ln(1-u^2)^2-\zeta_3-\frac{\pi^2}{6}\ln(1-u^2)\Big]\,,\\
p_1&=&-\frac{60-150u^2+50u^4+135u^6-122u^8+30^{10}}{240\sqrt3\,a^5(1-u^2)^3} -\frac{(1-u^2)(1+u^2)\ln(1-u^2)}{4\sqrt3\,a^5u^2}\,,\nn\\
p_2&=&\frac1{2419200\sqrt3\, a^9(1-u^2)^7}(184800-1554000u^2 +5381320u^4-10339420u^6\nn\\
&&+12585020u^8-10736670u^{10}+7061710u^{12}-3696335u^{14}+1395311u^{16}\nn\\
&&-309330u^2+28350u^2)\nn\\
&&\frac{(220-960u^2+1733u^4-1905u^6+1581u^8-861u^{10}+180u^{12}) \ln(1-u^2)}{2880 \sqrt3\,a^9u^2(1-u^2)^3}\nn\\
&&+\frac{(1-u^2)(1+u^2)\ln(1-u^2)^2}{24\sqrt3\,a^9u^2}\nn\\
&&+\frac{u^2(3-2u^2+u^4)}{24\sqrt3\,a^9(1-u^2)^2}Li_2(u^2)\,,\nn\\
p_3&=&-\frac{1}{9580032000\sqrt3\,a^{13}(1-u^2)^{11}}(1376723040\nn\\
&&-14743436400u^2+72819404240u^4-217955007160u^6\nn\\
&&+435604410648u^8-601596226704u^{10}+569248082304u^{12}\nn\\
&&-342331884048u^{14}+89935070610u^{16}+41112834477u^{18}\nn\\
&&-53144469816u^{20}+25843335646u^{22}-7385646827u^{24}\nn\\
&&+1379096730u^{26}-172764900u^{28}+10602900u^{30})\nn\\
&&-\frac{1}{14515200\sqrt3\,a^{13}u^2(1-u^2)^7}(1632344\nn\\
&&-11581976u^2+36973504u^4-67974936u^6+74132144u^8\nn\\
&&-39321550u^{10}-8177783u^{12}+29008167u^{14}-21172437u^{16}\nn\\
&&+7937571u^{18}-1598310u^{20}+141750u^{22})\ln(1-u^2)\nn\\
&&+\frac1{8640\sqrt3\,a^{13}u^4(1-u^2)^3}(270-1262u^2 +2685u^4-3010u^6+2215u^8\nn\\
&&-1584u^{10}+872u^{12} -180u^{14})\ln(1-u^2)^2\nn\\
&&-\frac{(1-u^2)(1+u^2)\ln(1-u^2)^3}{216\sqrt3\,a^{13}u^2}\nn\\
&&+\frac1{8640\sqrt3\,a^{13}(1-u^2)^6}(300-1980u^2+5300u^4 -7095u^6+4309u^8-5u^{10}\nn\\
&&-1389u^{12} +638u^{14}-90u^{16})Li_2(u^2)\nn\\
&&-\frac{5(1-2u^2+6u^4-2u^6+u^8)\ln(1-u^2)Li_2(u^2)}{144\sqrt3\, a^{13}u^2(1-u^2)^2}\nn\\
&&-\frac{5-10u^2+18u^4-2u^6+u^8}{72\sqrt3\,a^{13}u^2(1-u^2)^2} \Big[Li_3(1-u^2)\nn\\
&&\qquad\qquad+\ln(u)\ln(1-u^2)^2 -\zeta_3 -\frac{\pi^2}{6}\ln(1-u^2)\Big]\,.\eea

\end{document}